\documentclass[12pt]{article}

\newlength{\uppermar}
\newlength{\lowermar}
\newlength{\leftmar}
\newlength{\rightmar}

\usepackage[centertags]{amsmath}
\usepackage{amsfonts}
\usepackage{amssymb}
\usepackage{amsthm}
\usepackage[russian,english]{babel}
\usepackage{calc}

\title{Quantization of Donaldson-Uhlenbeck-Yau theory}
\author{S.L. Lyakhovich and  A.A. Sharapov}
\date{\it Department of Quantum Field Theory, Tomsk State University,
\\
Tomsk 634050, Russia \\
e-mail: {\rm sll@phys.tsu.ru, sharapov@phys.tsu.ru}}

\begin{document}

\maketitle

\begin{abstract}
A covariant path-integral quantization is proposed for the
non-Lagrangian gauge theory described by the
Donaldson-Uhlenbeck-Yau equation. The corresponding partition
function is shown to admit a nice path-integral representation in
terms of the gauged G/G K\"ahler WZW model. A relationship with
the $J$-formulation of the anti-self-dual Yang-Mills theory is
explored.

\end{abstract}

\section{Introduction}
A considerable number of fundamental field-theoretical models do
not admit any natural Lagrangian formulation. The list of examples
includes self-dual YM fields, chiral bosons, higher-spin gauge
theories, Siberg-Witten equations, superconformal theories with
extended supersymmetry, etc. The absence of a Lagrangian poses a
subtle question about a partition function one should use to
path-integral quantize a classical theory defined solely by
equations of motion. Obviously, this question cannot be answered
in intrinsic terms, one or another extra structure is needed over
and above the equations of motion. In the Lagrangian case, it is
the action $S$ that plays the role of such a structure and the
corresponding partition function is taken to be the Feynman
probability amplitude $\Psi= e^{\frac i\hbar S}$.

In our recent papers \cite{KLS}, \cite{LS3}, \cite{LS2}, a new
concept of a \textit{Lagrange anchor} was introduced with the aim
to extend the usual path-integral quantization technique beyond
the scope of Lagrangian mechanics. As a rough guide, the Lagrange
anchor is a geometric structure on the configuration space of
fields that allows one to construct a generalized Schwinger-Dyson
equation for the partition function $\Psi$ under far less
restrictive assumptions than the existence of action. Given a
partition function and a suitable integration measure, one can
compute the quantum averages of physical observables that
suffices, in principle, to reconstruct the whole quantum
theory\footnote{Notice that the notion of a Lagrange anchor has a
phase space counterpart allowing one to define an associative
$\ast$-product and quantum equations of motion on the space of
physical observables \cite{LS1}.}. In \cite{LS2}, the proposed
quantization scheme was exemplified by quantizing the Maxwell
electrodynamics with electric and magnetic currents and chiral
bosons in various dimensions.

In the present paper, we apply this general method to quantize the
nonlinear gauge theory whose classical dynamics are governed by
the Donaldson-Uhlenbeck-Yau (DUY) equation \cite{D}, \cite{UY}.
The DUY equation plays a prominent role both in physics and
mathematics. On the mathematics side, it provides a
differential-geometric tool for describing the moduli space of
stable holomorphic vector bundles over a K\"ahler manifold, the
problem that is of considerable interest to algebraic geometers.
In physics, this equation is of critical importance for the
paradigm of heterotic string compactification on a Calabi-Yau
manifold \cite{GSW}. Also notice the fact that in four dimensions,
the DUY equation reduces to the anti-self-dual Yang-Mills
equation.

An \textit{ad hoc} method for quantizing the DUY theory was
proposed some while ago by  Nair and Schiff \cite{NS}. The key
step of their approach is the reformulation of the DUY theory as a
Lagrangian topological field theory in one higher dimension, the
so-called K\"ahler-Chern-Simons (KCS) theory. The quantum
reduction by the first and second class constraints arising in the
KCS theory induces then a covariant quantization of the original
DUY theory. The work \cite{NS} also reveals  a deep interrelation
between the DUY theory and higher-dimensional counterparts of the
WZW model. In the present paper, the DUY model is quantized by the
systematic method proposed in \cite{KLS},\cite{LS3},\cite{LS2} for
general non-Lagrangian gauge theories. Although technically and
methodologically our approach is significantly different from that
by Nair and Schiff, the final results seem to agree with each
other. In particular, we derive a path-integral representation for
the partition function of the quantum DUY theory in terms of a
gauged $G/G$ WZW-like model on a K\"ahler manifold and establish
its connection with the so-called $J$-formulation of the DUY
theory proposed in \cite{NS}.

\section{A generalized Schwinger-Dyson equation}

In this section, we give a short and simplified presentation of
general quantization method \cite{LS3}, \cite{LS2} in the form
convenient for applying to the Yang-Mills type models.

In the covariant formulation of quantum field theory one usually
studies the path integrals of the form
\begin{equation}\label{PI}
    \langle \mathcal{O}\rangle =\int [d\varphi]
    \,\mathcal{O}\,e^{\frac i{\hbar}S}\,.
\end{equation}
After normalization, the integral defines the quantum average of
an observable $\mathcal{O}[\varphi]$ in the theory with action
$S[\varphi]$.  It is believed that evaluating the path integral
(\ref{PI}) for various reasonable observables $\mathcal{O}$, one
can extract all the physically relevant information about the
quantum dynamics of the model.

The partition function $\Psi [\varphi]=e^{\frac i{\hbar}S}$,
weighting the contribution of a particular field configuration
$\varphi$ to the quantum average, is known as the Feynman
probability amplitude. This amplitude can be defined as a unique
(up to a normalization factor) solution to the Schwinger-Dyson
(SD) equation
\begin{equation}\label{SDf}
    \left (\frac{\partial S}{\partial\varphi^i}+i\hbar\frac{\partial}{\partial
    \varphi^i}\right) \Psi[\varphi]=0\,.
\end{equation}
Performing the Fourier transform from the fields $\varphi$ to
their sources $J$, we can bring (\ref{SDf}) to a more familiar
form
\begin{equation}\label{SD}
\left( \frac{\partial
S}{\partial\varphi^i}(\hat\varphi)-J_i\right)Z[J]=0\,,\qquad
\hat\varphi{}^i\equiv i\hbar\frac{\partial}{\partial J_i}\,,
\end{equation}
where
\begin{equation}\label{Z}
    Z[J]=\int [d\varphi] e^{\frac{i}{\hbar}(S-J\varphi)}
\end{equation}
is the generating functional of Green's functions.

To guess how the Schwinger-Dyson equation could be generalized to
a theory whose classical equations of motion do not admit a
variational formulation, it might be instructive to start with the
following simple observations:

\vspace{2mm}

    $(i)$ Although the Feynman probability amplitude involves an action functional, the SD equation
    contains solely the equations of motion, not the action as such.

\vspace{2mm}

    $(ii)$ In the classical limit $\hbar\rightarrow 0$, the second
    term in the SD equation (\ref{SDf}) vanishes and the
    Feynman probability amplitude $\Psi$ turns to the Dirac distribution supported
    at the classical solutions to the field equations. Formally, $\Psi[\varphi]|_{\hbar\rightarrow 0} \sim
    \delta[\partial_i
    S]$ and one can think of the last expression as the
    \textit{classical partition function} \cite{Gozzietal}.

\vspace{2mm}

    $(iii)$ It is quite natural to treat the sources $J$ as the momenta
    canonically conjugate  to the fields $\varphi$, so that the only non-vanishing Poisson brackets
    are $\{\varphi^i,J_j\}=\delta^i_j$. Then one can regard the SD
    operators as resulting from the canonical quantization of the
    first class constraints $\Theta_i=\partial_i S-J_i\approx 0$
    on the phase space of fields and sources. Upon this
    interpretation, the Feynman probability amplitude describes  a unique physical state
    of a first-class constrained theory.
    This state is unique as  the ``number'' of the first class constraints $\Theta_i$ equals the
    ``dimension''
    of the configuration space of fields. Quantizing the constrained system in
    the   momentum representation yields  the SD equation (\ref{SD}) for
    the generating functional of Green's functions.

\vspace{2mm}

The above interpretation of the SD equations as operator first
class constraints on a physical wave-function suggests a direct
way to their generalization. Namely, consider a set of field
equations
\begin{equation}\label{T-eq}
T_a(\varphi^i)=0\,,
\end{equation}
which do not necessarily come from the variational principle. In
this case the (discrete parts of) superindices $a$ and $i$ may run
over completely different sets. Proceeding from the heuristic
arguments above, we can take the following ansatz for the $\varphi
J$-symbols of the Schwinger-Dyson operators:
\begin{equation}\label{TT}
\Theta_a=T_a(\varphi)-V_a^i(\varphi)J_i+ O(J^2)\,.
\end{equation}
The symbols are defined as formal power series in momenta
(sources) $J$ with leading terms being the classical equations of
motion. Requiring the Hamiltonian constraints $\Theta_a\approx 0$
to be first class, i.e.,
\begin{equation}\label{inv}
    \{\Theta_a, \Theta_b\}=U_{ab}^c \Theta_c \,,\qquad
    U_{ab}^c(\varphi,J)=C^c_{ab}(\varphi)+ O(J)\,,
\end{equation}
we obtain an infinite set of relations on the expansion
coefficients of $\Theta_a$. In particular, examining the
involution relations (\ref{inv}) to the leading order in $J$, we
find
\begin{equation}\label{anchor}
    V_a^i\partial_iT_b=G_{ab}+C_{ab}^c T_c
\end{equation}
for some structure functions
\begin{equation}\label{}
        G_{ab}(\varphi)=G_{ba}(\varphi)\,, \qquad
        C_{ab}^c(\varphi)=-C^c_{ba}(\varphi)\,.
\end{equation}
The value $V_a^i(\varphi)$, being defined by relation
(\ref{anchor}), is called the \textit{{Lagrange anchor}}. Under
the standard regularity conditions on the field equations
(\ref{T-eq}), any first order solution to (\ref{inv}), determined
by the Lagrange anchor $V$, has a prolongation to all orders in
$J$ \cite{LS3}. The symmetric matrix $G_{ab}$ is called the
generalized Van Vleck matrix.

For variational field equations, $T_a=\partial_i S$, one can set
the Lagrange anchor to be the unit matrix $V_a^i=\delta^i_a$. This
choice results in the standard Schwinger-Dyson operators
(\ref{SDf}, \ref{SD}) obeying the abelian involution relations.
Generally, the Lagrange anchor may be field-dependent and/or
noninvertible. If the Lagrange anchor is invertible, in which case
the number of equations must coincide with the number of fields,
then the operator $V^{-1}$ plays the role of integrating
multiplier in the inverse problem of calculus of variations. So,
the existence of the invertible Lagrange anchor amounts to the
existence of action. The other extreme choice, $V=0$, is always
possible and corresponds to a pure classical probability amplitude
$\Psi[\varphi]=\delta[T_a(\varphi)]$ supported at classical
solutions. Any nonzero Lagrange anchor, be it invertible  or not,
yields a ``fuzzy'' partition function describing nontrivial
quantum fluctuations in the directions spanned by the vector
fields $V_a=V_a^i\partial_i$.

In the non-Lagrangian case, the constraints (\ref{TT}) are not
generally the whole story. The point is that the number of
(independent) field equations can happen to be less than the
dimension of the configuration space of fields. In that case, the
field equations (\ref{T-eq})  do not specify a unique solution
with prescribed boundary conditions or, stated differently, the
system enjoys a gauge symmetry  generated by some on-shell
integrable vector distribution
$R_\alpha=R_\alpha^i(\varphi)\partial_i$. To allow for the gauge
invariance at the quantum level, one has to introduce the
additional first class constraints in the phase space of fields
and sources
\begin{equation}\label{R}
    R_\alpha =R_\alpha^i(\varphi)J_i+O(J^2)\approx 0\,.
\end{equation}
The leading terms of these constraints coincide with the  $\varphi
J$-symbols of the gauge symmetry generators and the higher orders
in $J$ are determined from the requirement that the whole set of
constraints $\Theta_I=(T_a, R_\alpha)$ to be the first
class\footnote{For a Lagrangian gauge theory we have
$T_i=\partial_iS-J_i$ and $R_\alpha =-R^i_\alpha T_i=R_\alpha^i
J_i$. In this case, one may omit the ``gauge'' constraints
$R_\alpha\approx 0$ as they are given by linear combinations of
the ``dynamical'' constraints $T_i\approx 0$.}. With all the gauge
symmetries included, the constraint surface $\Theta_I\approx 0$ is
proved to be a Lagrangian submanifold in the phase space of fields
and sources and the gauge invariant partition function is defined
as a unique solution to the \textit{generalized SD equation}
\begin{equation}\label{SDE}
    \hat{\Theta}_I\Psi=0\,.
\end{equation}
The last formula is just the definition of a physical state in the
Dirac quantization method \cite{HT}. A more systematic treatment
of the generalized SD equation within the BFV-BRST formalism can
be found in \cite{KLS}, \cite{LS3}.

In practice, it can be a problem to explicitly derive the
probability amplitude from the SD equation (\ref{SDE}), especially
in nonlinear field theories. In many interesting cases the
amplitude $\Psi[\varphi]$ is given by an essentially nonlocal
functional. More precisely, it can be impossible to represent
$\Psi$ as a (smooth) function of any local functional of fields
(by analogy with the Feynman probability amplitude $e^{\frac
i\hbar S}$ in a local theory with action $S$) even though the SD
equations (\ref{SDE}) are local. Fortunately, whatever the field
equations and Lagrange anchor may be, it is always possible to
write down a path-integral representation for $\Psi$ in terms of
some enveloping \textit{Lagrangian} theory. By now, two such
representations are known. The first one, proposed in \cite{KLS},
exploits the equivalence between the original dynamical system
described by the classical equations of motion $T_a=0$ and the
Lagrangian theory with action
\begin{equation}\label{top}
S[\varphi, J, \lambda]=\int_0^1 dt(\dot \varphi^i
J_i-\lambda^a\Theta_a )\,.
\end{equation}
The latter can be regarded as a Hamiltonian action of topological
field theory on the space-time with one more (compact) dimension
$t\in [0,1]$. The solution to the SD equation (\ref{SDE}) can be
formally represented by the path integral
\begin{equation}\label{top-PI}
    \Psi[\varphi_1] =\int [d\varphi][dJ][d\lambda] e^{\frac i\hbar
    S[\varphi,J,\lambda]}\,,
\end{equation}
where the sum runs over all trajectories with  $
\varphi(1)=\varphi_1$ and $J(0)=J(1)=0$. In \cite{LS2}, we used
such a representation  to perform a covariant quantization of the
chiral bosons in $d=4n+2$ dimensions in terms of  the
$(4n+3)$-dimensional  Chern-Simons theory.

An alternative approach to constructing a path-integral
representation for $\Psi$ is the \textit{augmentation method}
\cite{LS2}. With this method, one  augments the original
configuration space of fields $\varphi^i$ with the new fields
$\xi^a$, called the augmentation fields, and defines the action
\begin{equation}\label{S-aug}
S_{\mathrm{aug}}[\varphi, \xi]=
\xi^aT_a(\varphi)+G_{ab}(\varphi)\xi^a\xi^b+O(\xi^3)\,,
\end{equation}
where $G_{ab}$ is given by (\ref{anchor}), and the higher orders
in $\xi$ are determined from the condition that the (partially
averaged) amplitude
\begin{equation}\label{Psi-aug}
    \Psi[\varphi]=\int[d\xi]e^{\frac i\hbar S_{\mathrm{aug}}[\varphi,\xi]}
\end{equation}
obeys the SD equation (\ref{SDE}). There is also a simple
recursive algorithm allowing one to reconstruct (\ref{S-aug}) up
to any order in $\xi$'s \cite{LS2}. Notice that unlike the
topological model (\ref{top}), the augmented theory (\ref{S-aug})
is \textit{not} classically equivalent to the original
(non-)Lagrangian theory. So, the augmentation fields should not be
confused with a somewhat similar concept of ``auxiliary fields''
\cite{H}. With the amplitude (\ref{Psi-aug}), the quantum average
of an observable $\mathcal{O}$ can be written as
\begin{equation}\label{average}
    \langle \mathcal{O}\rangle =\int[d\varphi]
    \mathcal{O}[\varphi]\Psi[\varphi]=\int[d\varphi]
    [d\xi]\mathcal{O}[\varphi]e^{\frac i\hbar
    S_{\mathrm{aug}}[\varphi,\xi]}\,.
\end{equation}
It is significant that the action $S_{\mathrm{aug}}$ is given by a
local functional whenever the  Lagrange anchor and the equations
of motion are local. In that case, the integral (\ref{average}) is
similar in structure to (\ref{PI}), so the usual field-theoretical
tools of the Lagrangian theory can be still applied to evaluate
the quantum averages.

\section{Lagrange anchor for DUY theory}

Let $\mathcal{E}\rightarrow M$ be a holomorphic $G$-vector bundle
over a $2n$-dimensional K\"ahler manifold $M$ with the K\"ahler
2-form $\omega$. We take $G$ to be a compact Lie group and denote
by $\mathcal{G}$ its Lie algebra. Consider a linear connection $A$
on $\mathcal{E}$. As any of the 2-forms on a complex manifold, the
curvature $F$ of the connection is decomposed into the sum of the
$\mathcal{G}$-valued $(2,0)$, $(0,2)$, and $(1, 1)$-forms on $M$.
The Donaldson-Uhlenbeck-Yau equations read
\begin{equation}\label{DUY1}
    F^{(2,0)}=0\,,\qquad F^{(0,2)}=0\,,
    \end{equation}
    \begin{equation}\label{DUY2}
    \omega^{n-1}\wedge F^{(1,1)}=0\,.
\end{equation}
The first two equations just mean that the connection is
holomorphic and the last condition is equivalent to the stability
of the holomorphic vector bundle in algebraic geometry \cite{D},
\cite{UY}.

For $n=1$ equations (\ref{DUY1}, \ref{DUY2}) reduce to a single
zero curvature condition $F=0$. Setting $n=2$ one obtains three
independent equations that are equivalent to the anti-self-duality
condition for the curvature 2-form $F$.  Since the solutions to
the DUY equations constitute a part of solutions to the
corresponding Yang-Mills equations in any dimension, one may
regard (\ref{DUY1}, \ref{DUY2}) as a higher-dimensional
generalization of the anti-self-dual YM theory in four dimensions.

Note that equations (\ref{DUY1}, \ref{DUY2}), being gauge
invariant, are linearly independent, so no Noether identities are
possible. This property is a particular manifestation of a
non-Lagrangian nature of the DUY equations. In a Lagrangian
theory, any gauge symmetry gives rise to a Noether identity and
vice versa. Although the DUY equations are not Lagrangian, they
admit a good Lagrange anchor that leads, as we will see, to a
reasonable quantum theory.

Denote by $\mathcal{A}$ the affine space of all connections on
$\mathcal{E}$. Locally, any connection on $\mathcal{E}$ is
represented by a pair of $(1,0)$ and $(0,1)$-forms $(A, \bar A)$
valued in the Lie algebra $\mathcal{G}$. In terms of the gauge
potentials $A$ and $\bar A$, the homogeneous components of the
curvature $F$ read
\begin{equation}\label{}
\begin{array}{c}
F^{(2,0)}=\partial A + A\wedge A\,,\qquad  F^{(0,2)}=\bar\partial
\bar A+
\bar A\wedge \bar A\,,\\[5mm]
F^{(1,1)}=\partial \bar A+\bar
\partial A
    +A\wedge \bar A+\bar A\wedge A\,,
    \end{array}
\end{equation}
where $\partial$ and $\bar \partial$ are holomorphic and
anti-holomorphic parts of the de Rham differential $d$. The
canonical symplectic structure on the cotangent bundle of
$\mathcal{A}$ reads
\begin{equation}\label{}
    \Omega =\int_M \mathrm{Tr}(\delta A\wedge \delta P)
    +\int_M\mathrm{Tr}(\delta\bar A\wedge \delta\bar P)\,,
\end{equation}
with $P$ and $\bar P$ being, respectively,  $(n-1,n)$ and
$(n,n-1)$-forms on $M$ with values in $\mathcal{G}$. As we have
explained in Sec.2, one may regard the fields $P$ and $\bar P$,
playing the role of canonical momenta, as the sources to the gauge
fields $A$ and $\bar A$. Following the general prescription of
Sec.2, we introduce the corresponding set of first class
constraints (\ref{TT}) on the phase space of fields and sources:
\begin{equation}\label{constr}
\begin{array}{c}
    T_0 = \omega^{n-1}\wedge F^{(1,1)}+k(DP-\bar D\bar P)\approx 0\,,\\[5mm]
T_+=F^{(2,0)}\approx 0\,,\qquad T_-=F^{(0,2)}\approx 0\,,\\[5mm]
    R=DP+\bar D\bar P\approx 0\,.
    \end{array}
\end{equation}
Here $D$ and $\bar D$ are the covariant differentials  associated
with the gauge fields $A$ and $\bar A$ and $k$ is a complex
parameter.

Let us comment on the structure of the constraints (\ref{constr}).
The constraint $T_0\approx 0$ is just a one-parameter deformation
of the classical stability condition (\ref{DUY2}) by the momenta
dependent term. According to our terminology, this term defines
(and is defined by) a Langrange anchor compatible with the
classical equations of motion (\ref{DUY1}, \ref{DUY2}).  The rest
of the DUY equations, namely the holomorphy conditions
(\ref{DUY1}), remain intact and define the holonomic constraints
$T_\pm\approx 0$ on the phase space of fields and sources. In
physical terms, this means that the quantum fluctuations are
nontrivial only for that part of classical dynamics which is
governed by the stability condition. Finally, the constraint $R$
reflects the presence of gauge symmetries. The Hamiltonian action
of $R$ induces the standard gauge transformations on the
configuration space of fields $\mathcal{A}$. Taken together, the
Hamiltonian constraints (\ref{constr}) define a topological field
theory (\ref{top}) on the cotangent bundle of $\mathcal{A}$, which
is found to be classically equivalent to the original
non-Lagrangian dynamics (\ref{DUY1}, \ref{DUY2}).

To describe the Poisson algebra of the first class constraints,
it is convenient to interpret them as linear functionals (de
Rham's currents) on an appropriate  space of $\mathcal{G}$-valued
forms. Define
\begin{equation}\label{constr1}
T_0(\varepsilon_0)=\int_M\mathrm{Tr}(\varepsilon_0 \wedge
T_0)\,,\qquad T_{\pm}(\varepsilon_{\pm})=\int_M\mathrm{Tr}
(\varepsilon_{\pm}\wedge T_{\pm})\,,\qquad R(\varepsilon\,)=\int_M
\mathrm{Tr}(\varepsilon\wedge R)\,,
\end{equation}
where $\varepsilon_0$, $\varepsilon_\pm$, and $\varepsilon$ are
gauge parameters whose form degrees are complementary to the
degrees of corresponding constraints.  The Poisson brackets of the
constraints read
\begin{equation}\label{PB}
    \begin{array}{ll}
\{T_0(\varepsilon),T_0(\varepsilon')\}=k^2R([\varepsilon,\varepsilon'])\,,\qquad&
\{R\,(\varepsilon),R\,(\varepsilon')\}=R\,([\varepsilon,\varepsilon'])\,,
\\[5mm]
\{R(\varepsilon),T_0(\varepsilon')\}=T_0([\varepsilon,\varepsilon'])\,,\qquad&
\{R(\varepsilon),T_{\pm}(\varepsilon')\}=T_\pm([\varepsilon,\varepsilon'])\,,\\[5mm]
\{T_0(\varepsilon),T_\pm(\varepsilon')\}=\pm
kT_\pm([\varepsilon,\varepsilon'])\,,\qquad&\{T_\pm(\varepsilon),T_\pm(\varepsilon')\}=0\,.
    \end{array}
\end{equation}
Upon canonical quantization the first class constraints
(\ref{constr}) turn to the Scwinger-Dyson
operators\footnote{Hereinafter we set $\hbar =1$.}

\begin{equation}\label{q-constr1}
    \hat{T}_0=\omega^{n-1}\wedge F^{(1,1)} -i k \left(D\frac{\delta}{\delta
    A}- \bar D\frac{\delta}{\delta \bar A}\right)\,,\qquad \hat R=
    -i\left(D\frac{\delta}{\delta A}+\bar D\frac{\delta}{\delta \bar A}
    \right)\,,
\end{equation}

\begin{equation}\label{q-constr2}
    \hat{T}_+=F^{(2,0)}\,,\qquad \hat{T}_-=F^{(0,2)}\,.
\end{equation}

The partition function $\Phi$ on the configuration space of fields
$\mathcal{A}$ is now defined as a unique (up to a multiplicative
constant) functional annihilated by all the operator constraints
(\ref{q-constr1}, \ref{q-constr2}). Imposing the operators
 of holonomic constraints (\ref{q-constr2}) yields the following
expression for the  partition function:
\begin{equation}\label{Phi}
    \Phi = \delta[T_+] \delta[T_-]\Psi\,,
\end{equation}
where the function $\Psi[A,\bar A]$ is annihilated by the
residuary constraints   (\ref{q-constr1}),
\begin{equation}\label{TPsi}
\hat{R}\Psi=0\,,\qquad \hat{T}_0\Psi=0\,.
\end{equation}
The first equation just says that $\Psi$, and hence $\Phi$, are
gauge invariant functionals of $A$ and $\bar A$. The second
equation is the quantum counterpart of the stability condition
(\ref{DUY2}); being nonlinear, it is the most challenging equation
to solve. One of the complications in  solving (\ref{TPsi}) is
that there is no way to represent $\Psi$ as a function of any
local functional of the fields $A$ and $\bar A$. Nonetheless, by
making use the augmentation method, we can construct a
path-integral representation for $\Psi$ in terms of a local action
functional on an augmented configuration space. By definition, the
augmentation fields take values in the space dual to the space of
equations of motion. Therefore,  we extend the original
configuration space of fields $\mathcal{A}$ by introducing the set
of new fields $\xi=(\Lambda, \bar \Lambda, B)$, where $\Lambda$
and $\bar \Lambda$ are $\mathcal{G}$-valued $(n-2,n)$ and
$(n,n-2)$-form fields, respectively, and $B$ is a
$\mathcal{G}$-valued scalar field on $M$. Then up two the first
order in $\xi$'s the action (\ref{S-aug}) reads
\begin{equation}\label{S-full}
S_{\mathrm{aug}} = S[A,\bar A, B]+\int_M \mathrm{Tr}(\Lambda
\wedge F^{(2,0)}+\bar \Lambda \wedge F^{(0,2)})\,,
\end{equation}
where
\begin{equation}\label{S}
    S[A,\bar A, B]=\int_M \omega^{n-1} \wedge \mathrm{Tr}(B
F^{(1,1)})+ O(B^2)\,.
\end{equation}
Note that the fields  $\Lambda$ and $\bar \Lambda$ enter the
action $S_{\mathrm{aug}}$ only linearly because the corresponding
constraints are holonomic.  The integration over these fields by
formula  (\ref{Psi-aug}) simply reproduces the delta-functions in
(\ref{Phi}). So we can focus our attention on the action
(\ref{S}).  Applying the general procedure from \cite{LS2} allows
one, in principle, to reconstruct (\ref{S}) up to any order in
$B$. As a practical matter, it is better to work in terms of the
group valued field $g=e^B$ rather than the Lie algebra valued
field $B$. With the field $g$, we are able to present a closed
expression for (\ref{S}), which appears to be nothing but a gauged
version of the K\"ahler WZW model \cite{NS}, \cite{IKU} (KWZW
model for short). The details of the construction are exposed in
the next section.

\section{The DUY equation and gauged G/G KWZW model}

Our staring point is the action of the KWZW model associated with
the K\"ahler manifold $(M,\omega)$ and the Lie group $G$. The
basic field of the model is a smooth mapping $g: M\rightarrow G$
and the action is
\begin{equation}\label{KWZW}
    S_0[g]=\frac12 \int_M \omega^{n-1}\wedge \mathrm{Tr}(g^{-1}\partial
g\wedge g^{-1}\bar\partial g) -\frac16 \int_{M \times
I}\tilde{\omega}^{n-1}\wedge
\mathrm{Tr}(\tilde{g}^{-1}d\tilde{g}\wedge
\tilde{g}^{-1}d\tilde{g}\wedge \tilde{g}^{-1}d\tilde{g})\,.
\end{equation}
In this expression, $I=[0,1]\subset \mathbb{R}$, $\tilde{g}$
denotes an extension of $g$ to the product manifold $M\times I$,
and $\tilde{\omega}$ is the pull-back of the K\"ahler form
$\omega$ with respect to the canonical projection $M\times
I\rightarrow M$. More precisely, we identify one boundary
component of $M\times I$, say $M\times \{1\}$, with the original
space $M$ and extend $g$ to $M\times I$ in such a way that it
tends to some fixed field $g_0$ on the other component of the
boundary $M\times \{0\}$; in so doing, the field $g$  falls into
the same homotopy class as $g_0$. Hereafter we assume that $g$
takes values in any unitary irreducible representation of $G$ and
$\mathrm{Tr}$ is the ordinary matrix trace.

In case $n=1$ the action (\ref{KWZW}) reduces to the familiar
action of  the WZW model \cite{W}. Like its two-dimensional
prototype, the higher-dimensional KWZW model enjoys the
infinite-dimensional symmetry
\begin{equation}
g\rightarrow h_1 g \bar h_2,
\end{equation}
$h_{1,2}$  being holomorphic mappings  to $G$, which can be easily
seen from the $2n$-dimensional analog of the Polyakov-Wiegmann
formula
\begin{equation}\label{PW}
S_0[gh]=S_0[g]+S_0[h]+\int_M\omega^{n-1}\wedge
\mathrm{Tr}(g^{-1}\partial g\wedge \bar\partial hh^{-1})\,.
\end{equation}
In particular, the model is invariant under the adjoint action of
$G$: $g\rightarrow hgh^{-1}$, $\forall h\in G$.

 The KWZW model possesses many other interesting properties and,
 as we show below, it results as a part of the augmented action for the DUY model.
Recall that in our treatment the field $g$ is identified with the
augmentation field for the non-Lagrangian equation (\ref{DUY2}).
The total action (\ref{S}) is obtained from (\ref{KWZW}) by
``gauging'' the adjoint action of $G$ through introduction of the
minimal coupling with the gauge fields $(A,\bar A)$. The
construction is patterned after the much studied two-dimensional
case \cite{GK} and results in the following action for the gauged
G/G KWZW model:
\begin{equation}\label{G-KWZW}
S[g,A,\bar A]=S_0[g] + \int_M \omega^{n-1}\wedge \mathrm{Tr}(
A\wedge \bar\partial g g^{-1}- g^{-1}\partial g\wedge \bar A
+A\wedge \bar A- A\wedge g\bar A g^{-1})\,.
\end{equation}
One can easily see that the action is invariant under the
infinitesimal gauge transformations
\begin{equation}\label{}
    \delta g=[u,g]\,,\qquad \delta A =-Du=-\partial
    u-[A,u]\,,\qquad \delta\bar A=-\bar D u=-\bar\partial u-[\bar
    A,u]\,.
\end{equation}
The total variation of (\ref{G-KWZW}) is given by
\begin{equation}\label{}
\begin{array}{c}
\delta S=\delta_g S+\delta_A S +\delta_{\bar A}S \,,\\[5mm]
\displaystyle \delta_g S =\int_M\omega^{n-1}\wedge
\mathrm{Tr}\left[ g^{-1}\delta g (F^{(1,1)}+\bar D(g^{-1}Dg))
\right]= \int_M\omega^{n-1}\wedge \mathrm{Tr}\left[  \delta
gg^{-1}(F^{(1,1)}-D(\bar Dg g^{-1}))\right],\\[7mm]
\displaystyle \delta_AS=\int_M\omega^{n-1} \wedge
\mathrm{Tr}(\delta A\wedge \bar Dgg^{-1})\,,\qquad \delta_{\bar
A}S=\int_M\omega^{n-1} \wedge \mathrm{Tr}(\delta\bar A\wedge
g^{-1} Dg)\,.
\end{array}
\end{equation}
So the classical equations of motion can be written as
\begin{equation}\label{}
    \omega^{n-1}\wedge F^{(1,1)}=0\,,\qquad Dg=0\,, \qquad \bar D g=0\,.
\end{equation}
We see that the dynamics of the fields $A$ and $\bar A$ are
completely decoupled from the dynamics of the augmentation field
$g$ and are governed by the DUY equation (\ref{DUY2}). Such a
structure of equations is typical for an augmented theory
\cite{LS2}.

Now we claim  that the path integral\footnote{In accordance with
our definition of the KWZW action (\ref{KWZW}), the sum runs over
all fields belonging to a fixed homotopy class $[g_0]$.}
\begin{equation}\label{Psi}
    \Psi[A,\bar A]=\int [dg]\,e^{\frac i{2k}S[g,A,\bar A]}\,,
\end{equation}
where $[dg]$ is induced by the Haar measure on $G$, yields a
desired solution to the Schwinger-Dyson equations (\ref{TPsi}).
The statement is proved simply by substituting (\ref{Psi}) into
(\ref{TPsi}) and differentiating under the integral sign. We have
\begin{equation}\label{check}
\begin{array}{c}
\displaystyle \hat{T}_0\Psi =\frac12\int_M [dg]\, \omega\wedge
[2F^{(1,1)}+
D(\bar Dgg^{-1})-\bar D(g^{-1}Dg)]e^{\frac i{2k}S[g,A\bar A]}\\[5mm]
\displaystyle =\frac12\int_M[dg]\, (V_L +V_R)e^{\frac
i{2k}S\,[g,A,\bar A]}\,.
\end{array}
\end{equation}
Here $V_L=t_aV^a_L$ and $V_R=t_aV^a_R$ are the first-order
variational operators associated with the basis $\{V^a_L\}$ and
$\{V^a_R\}$ of the left- and right-invariant vector fields on the
gauge group $\hat G$ (the group of maps from $M$ to $G$). These
vector fields are completely specified by the relations
\begin{equation}\label{}
    i_{V^a_L}(g^{-1}\delta g)=i_{V^a_R}(\delta
    gg^{-1})=t_a\,,
\end{equation}
where $\{t_a\}$ are the generators of the Lie algebra
$\mathcal{G}$ with $\mathrm{Tr}(t_at_b)=\delta_{ab}$. Since the
integration measure $[dg]$ is formally invariant under the action
of $V^a_L$ and $V^a_R$, we deduce that the integrand in
(\ref{check}) is a total divergence. Assuming that one can
integrate by parts in functional space, the right-hand side of
(\ref{check}) vanishes. Although the gauge invariance of the
amplitude $\Psi$ is obvious, it is instructive to verify it
directly:
\begin{equation}\label{check-up}
\hat{R} \,\Psi=\frac1{2k}\int_M [dg]\,(V_L-V_R)e^{\frac
i{2k}S\,[g,A,\bar A]}=0\,,
\end{equation}
by the same reasons as above.

Given the partition function (\ref{Phi}, \ref{Psi}), the vacuum
expectation value of a gauge invariant observable
$\mathcal{O}[A,\bar A]$ is defined by
\begin{equation}\label{q-av}
    \langle \mathcal{O}\rangle = \frac1{\mathrm{vol} (\hat G)} \int [dg][dA][d\bar
    A]\mathcal{O}\,\delta[F^{(2,0)}]\delta[F^{(0,2)}]\,e^{\frac i{2k}S[g,A,\bar
    A]}\,,
\end{equation}
where $[dA][d\bar A]$ is the translation-invariant measure on the
space of all connections $\mathcal{A}$. Since $\Psi$ is gauge
invariant it is natural to divide by the volume of the gauge group
$\hat G$. A more rigor treatment of the integral within the BV
formalism involves the standard gauge-fixing procedure \cite{HT}.
The expression (\ref{q-av}) is in a sense final if not particulary
convenient for perturbative calculations because of delta-function
factors in the integrand. To bring the path integral (\ref{q-av})
into the usual form (\ref{PI}) one can either replace the action
$S[g, A,\bar A]$ with (\ref{S-full}) and extend integration over
the Lagrange multipliers $\Lambda$, $\bar \Lambda$, or directly
solve the holonomic constraints $T_{\pm}\approx 0$ in terms of
some unconstrained fields. Observe that locally any solution to
(\ref{DUY1}) is representable in the form
\begin{equation}\label{sol}
A=h^{-1}\partial h\,,\qquad \bar A=-\bar \partial
h^\dagger(h^\dagger)^{-1}
\end{equation}
for some $G^{\mathbb{C}}$-valued field $h$. And vice versa,  for
any $h\in\hat{G}^\mathbb{C}$, the gauge potentials (\ref{sol})
satisfy the holomorphy conditions (\ref{DUY1}). The representation
(\ref{sol}) goes back to the work of Yang \cite{Y}, where it was
originally introduced in the context of anti-self-dual YM fields.
On substituting (\ref{sol}) into (\ref{G-KWZW}), we get the action
$S[g,h]$ which is the functional of the $G$-valued field $g$ and
$G^\mathbb{C}$-valued field $h$ and which is invariant under the
gauge transformations
\begin{equation}\label{g-t}
  g\rightarrow ugu^{-1}\,,\qquad   h \rightarrow hu^{-1} \,.
\end{equation}
Using the PW formula (\ref{PW}), we can write this action as the
difference of two explicitly gauge invariant terms
\begin{equation}\label{}
    S[g,h]=S_0[hgh^\dagger]-S_0[hh^\dagger]\,.
\end{equation}

Let $J=hh^\dagger$. The field $J$ takes values in
positive-definite Hermitian matrices. By making use the polar
decomposition of a nondegenerate matrix, we can write
$h=\sqrt{J}u$, where $\sqrt{J}$ is still Hermitian and
positive-definite matrix representing the points of the
homogeneous space $G^\mathbb{C}/G$, while $u\in G$ is unitary.
Then  the Haar measure on $\hat{G}^\mathbb{C }$ is factorized as
   $ [dh]=[du][dJ]$.
The integration over the unitary factor $u\in \hat{G}$ gives just
the volume of the gauge group, $\int [du]=\mathrm{vol}(\hat G)$.

Performing the change of variables (\ref{sol}), we can rewrite
(\ref{q-av}) as
\begin{equation}\label{q-av1}
\langle
{\mathcal{O}}\rangle=\frac{1}{\mathrm{vol}(\hat{G})}\int[dg][dh]\Delta[h]\mathcal{O}
    e^{\frac
    i{2k}(S_0[hgh^\dagger]-S_0[hh^\dagger])}
= \frac{C}{\mathrm{vol}(\hat{G})}\int[dh]\Delta[h]\mathcal{O}
    e^{-\frac i{2k}S_0[hh^\dagger]}
    \,,
    \end{equation}
where the local measure $\Delta'[h]$ is defined formally by the
relation
\begin{equation}
[dA][d\bar A]=[dF^{(2,0)}][dF^{(0,2)}][dh]\Delta[h]
\end{equation}
and
\begin{equation}\label{}
C=\int[dg]
    e^{\frac i{2k}S_0[hgh^\dagger]}=\int[dg]e^{\frac
    i{2k}S_0[g]}\,.
\end{equation}
(The last equality is just a formal extension to infinite
dimensions of the invariance of the Haar integral $\int_G dg f(h_1
g h_2)=\int_G dg f(g)$, where $f$ is analytic  on $G^{\mathbb{C}}$
and $h_1,h_2\in G^\mathbb{C}$.) In the absence of gauge anomalies,
the integrand of (\ref{q-av1}) is to be invariant under the gauge
transformation (\ref{g-t}) and it is reasonable to assume that
$\Delta =\Delta'[hh^\dagger]$ and $\mathcal{O}
=\mathcal{O}'[hh^\dagger]$. If $G$ is abelian, then $\Delta$ is
just an essential constant factor. By making use the polar
decomposition $h=\sqrt{J} u$, we finally obtain
  \begin{equation}
   \displaystyle  \langle
{\mathcal{O}}\rangle= \frac{C}{\mathrm{vol}(\hat{G})}\int
  [du][dJ]\Delta'[J]\mathcal{O}'[J]e^{-\frac i{2k}S_0[J]}=C\int
  [dJ]\Delta'[J] \mathcal{O}'[J] e^{-\frac i{2k}S_0[J]}\,.
\end{equation}
The last integral expresses the quantum average of a gauge
invariant observable $\mathcal{O}$ in terms of the KWZW-like
action $S_0[J]$ and the local measure $\Delta'[J]$ associated to
the homogeneous space $\hat{G}^\mathbb{C}/G$. By construction, the
field $J$ describes the (local) physical modes of the gauge fields
$A$ and $\bar A$. The extremum points  of the action $S_0[J]$ are
defined by the equation
\begin{equation}\label{J}
    \bar \partial (\omega^{n-1}\wedge J^{-1}\partial J)=0\,.
\end{equation}
We could also arrive at this equation  by simply substituting the
holomorphic potentials (\ref{sol}) into the DUY equation
(\ref{DUY2}). In the special case that $M$ is a four-dimensional
K\"ahler manifold, equation (\ref{J}) was intensively studied in
the past under the name of $J$-formulation for the anti-self-dual
YM theory (see e.g. \cite{BFNY}, \cite{CGW}, \cite{NS}).

\subsection*{Acknowledgments} We wish to thank Petr Kazinski for
fruitful  collaboration at the early stage of this work. The work
was partially supported by the RFBR grant 06-02-17352 and the
grant for Support of Russian Scientific Schools 1743.2003.2.


\begin{thebibliography}{BGHHW}

\bibitem{KLS} P.O. Kazinski, S.L. Lyakhovich and A.A. Sharapov,
\textit{Lagrange Structure and Quantization}, JHEP \textbf{07}
(2005) 076.

\bibitem{LS3} S.L. Lyakhovich and A.A. Sharapov, \textit{Schwinger-Dyson equation for non-Lagrangian field
theory}, JHEP \textbf{02} (2006) 007.

\bibitem{LS2} S.L. Lyakhovich and A.A. Sharapov, \textit{Quantizing non-Lagrangian gauge theories: an augmentation
method}, JHEP \textbf{01} (2007) 047.

\bibitem{LS1} S.L. Lyakhovich and A.A. Sharapov,
\textit{BRST theory without Hamiltonian and Lagrangian},  JHEP
\textbf{03} (2005) 011.


\bibitem{D} S.K. Donaldson, \textit{Anti-self-dual Yang-Mills connections over complex algebraic surfaces and stable vector bunldles},
 Proc. London Math. Soc. \textbf{50} (1985) 1.

\bibitem{UY} K.K. Uhlenbeck and S.T. Yau, \textit{On the existence of hermitian Yang-Mills connections in stable vector bundles},
 Commun. Pure Appl. Math. \textbf{39 }(1986) 257.



\bibitem{GSW} M.B. Green, J.H. Schwarz and E. Witten, Superstring
Theory, vol. 2 (Cambridge University Press, 1987).


\bibitem{NS} V.P. Nair and J. Schiff, \textit{ K\"ahler-Chern-Simons Theory and Quantization of Instanton Moduli Spaces},
Phys. Lett. \textbf{B246} (1990) 423;
\textit{K\"ahler-Chern-Simons Theory and Symmetries of
Anti-Self-Dual Gauge Fields},
 Nucl. Phys. \textbf{B371} (1992) 329.

\bibitem{Gozzietal} E.~Gozzi, \textit{Hidden BRS invariance in classical mechanics},
Phys. Lett. \textbf{B201}(1988) 525;  E.~Gozzi, M.~Reuter and
W.D.~Thacker, \textit{Hidden BRS Invariance in Classical
Mechanics. 2}, Phys. Rev. \textbf{D40} (1989) 3363.


\bibitem{HT}  M. Henneaux and C. Teitelboim, \textit{Quantization of Gauge
Systems} (Princeton U.P., NJ, 1992).


\bibitem{H} M. Henneaux, \textit{Elimination Of The Auxiliary Fields In The Antifield Formalism}, Phys. Lett. \textbf{B238} (1990)
299.


\bibitem{IKU} T. Inami, H. Kanno and T. Ueno, \textit{Higher dimensional WZW Model on K\"ahler Manifold and Toroidal Lie
Algebra}, Mod. Phys. Lett. \textbf{A12} (1997) 2757.


\bibitem{GK} K. Gawedzki and A. Kupiainen, \textit{G/H conformal field
theory from gauged WZW model}, Phys. Lett. \textbf{B215} (1988)
119; \textit{Coset construction from functional integrals}, Nucl.
Phys. \textbf{B320} (1989) 625.


\bibitem{W} E. Witten, \textit{Non-Abelian Bosonization in Two
Dimensions}, Commun. Math. Phys. \textbf{92} (1984) 455.


\bibitem{Y} C.N. Yang, \textit{Condition of Self-Duality for SU(2) Gauge Fields on Euclidean Four-Dimensional
Space}, Phys. Rev. Lett. \textbf{38} (1977) 1377.


\bibitem{BFNY}
Y. Brihaye, D.B. Fairlie, J. Nuyts and  R.G. Yates,
\textit{Properties of the sefldual equations for an SU(n) gauge
theory}, J. Math. Phys. \textbf{19} (1978) 2528.

\bibitem{CGW} L.L. Chau, M.-L. Ge and Y.-S. Wu, \textit{Kac-Moody algebra in the self-dual Yang-Mills
equation}, Phys. Rev. \textbf{D25} (1982) 1086.

\end{thebibliography}
\end{document}